\documentstyle[preprint,epsf,aps]{revtex}

\def\etal{{\it et al.}\thinspace}

\def\eg{{\it e.g.,}\thinspace}

\begin{document}
%
\title{\bf X-ray photoemission study of NiS$_{2-x}$Se$_x$ ($x$ = 0.0 - 1.2)}

\author{S.~R.~Krishnakumar \cite{ictp} and D.~D.~Sarma \cite{jnc}}

\address{Solid State and Structural  Chemistry Unit,
Indian Institute of Science, Bangalore 560012, India}

\maketitle
\begin{abstract}

Electronic structure of NiS$_{2-x}$Se$_x$ system has been
investigated for various compositions ($x$) using x-ray
photoemission spectroscopy. An analysis of the core level as well
as the valence band spectra of NiS$_2$ in conjunction with
many-body cluster calculations provides a quantitative description
of the electronic structure of this compound. With increasing Se
content, the on-site Coulomb correlation strength ($U$) does not
change, while the band width $W$ of the system increases, driving
the system from a covalent insulating state to a $pd$-metallic
state.

\end{abstract}

\vspace{0.2in}



\section{Introduction}

Pyrite type disulfides of 3$d$  transition metals have been
extensively studied as these systems  are expected to  be well
suited  for the experimental  investigations of electron
correlation effects  in narrow band electron systems.   In the
isostructural pyrite  series, $M$S$_2$ ($M$~=~Fe, Co, Ni, Cu  and
Zn), the physical properties of  the system evolves from the
progressive filling of the  3$d$ bands of $e_g$ symmetry and
exhibit a wide variety  of electrical and magnetic properties
\cite{wil@4,yoffe@4}.   The ground state metallic or insulating
properties of the series appear mostly  in agreement with the
single particle band theory. Thus,  FeS$_2$ and ZnS$_2$  are
insulators where the $e_g$ band  is entirely empty for FeS$_2$ and
totally occupied for ZnS$_2$, whereas CoS$_2$ and  CuS$_2$ with 1/4  and 3/4  band
fillings  are metals,  as expected \cite{wil@4,yoffe@4}. However,
NiS$_2$, in spite of its half-filled $e_g$ band, is  insulating in
contrast to results based on band structure calculations and is
thought to be  driven by electron correlations giving rise to the
Mott insulating state \cite{honigrev@4}.  It  is found that
NiSe$_2$, with the same pyrite  structure, is  completely miscible
with NiS$_2$  in  the entire composition  range,   forming  the
solid   solution NiS$_{2-x}$Se$_x$. NiSe$_2$ is metallic and thus
the ground state of the solid solution changes over  from
insulating  to a  metallic one at  $x_c\sim0.43$
\cite{sudojmmm@4},  without any change in  the symmetry  of  the
crystal structure \cite{endo@4} and is believed to be  an ideal
testing ground  for various  many-body theories  for
metal-insulator transitions  in strongly correlated narrow band
electron systems \cite{honigrev@4}.

At room temperature, NiS$_2$ is a paramagnetic insulator with a
band gap of 0.3~eV, estimated  from   optical studies
\cite{kautz@4}, while the transport measurements suggest an
activation energy of 0.2~eV \cite{honigrev@4}.  It is also found
that for  a narrow range of composition near $x_c$, the system
undergoes a transition from antiferromagnetic insulator  to
antiferromagnetic  metal with  decreasing  temperature
\cite{honigres@4}.    The  low temperature antiferromagnetic
metallic  state   vanishes  around $x=1.0$,  leading  to a
paramagnetic  metallic ground  state  of the system
\cite{honigres@4}. NiS$_2$ crystallizes in the  cubic pyrite
structure ($Pa\overline{3} $ space  group) with the lattice
parameter, $a=5.620$~\AA\ \cite{stru@4}. Each Ni atom  is
coordinated with 6 sulphur atoms in a slightly distorted
octahedral environment with the 3$d$  level split essentially into
a  lower lying $t_{2g}$ triplet  and a higher lying $e_g$ doublet.
The  Ni-S distance in NiS$_2$ ($d_{Ni-S}\simeq$~2.40~\AA)  is
slightly larger  than  that  in other divalent sulphides  of Ni,
such as NiS($d_{Ni-S}\simeq$~2.39~\AA) and BaNiS$_2$
($d_{Ni-S}\simeq$~2.32~\AA).   One    characteristic feature  of
the   structure of NiS$_2$ is the  presence  of very short S-S
bonds ($d_{S-S}\simeq$~2.06~\AA) compared to that      in other
sulphides, such as NiS ($d_{S-S}\simeq$~3.44~\AA)  and BaNiS$_2$
($d_{S-S}\simeq$~3.14~\AA). This  leads  to the formation  of
S$_2^{2-}$  dimers, indicating  the presence of strong S-S
interactions in the system.

The electronic structure of NiS$_{2-x}$Se$_x$ has been studied
extensively over the years, though mainly using UV-photoemission
spectroscopy. However, a recent high-resolution UV photoemission
study of this series with $x\le0.4$ \cite{nis2prl}  established
that the surface electronic structure of this system behaves  very
differently compared to that of the bulk. These differences can be
clearly observed with the surface sensitive UV photoemission
technique, appearing close to the Fermi energy and exhibiting
interesting changes in the electronic structure as a function of
the temperature. These changes can only be seen within 500 meV of
$E_F$ and do not appear to affect even the UV photoemission
spectrum in the main valence band region appearing nearly 2 eV
below $E_F$. X-ray photoemission spectroscopy is known to be more
bulk sensitive than UV-photoemission spectroscopy; therefore, it
is more appropriate to use this technique to study the gross
electronic structure of NiS$_2$ and related compounds and its
evolution across the solid solution series. It is also well known
that electron-correlation effects are important to describe the
electronic structure of these systems while single-particle band
theories  fail. Different configuration-interaction models
including electron correlation effects have been proposed to
explain the valence band (VB) \cite{fujivbcal@4} and core level
spectra \cite{fujicorecal@4,luigi@4} in the past; however, there
has been no consistency between different models used and also
between models used for the core level and valence band spectra.
In the present study, we investigate the electronic structure of
NiS$_{2-x}$Se$_x$ system using X-ray photoelectron spectroscopic
measurements in conjunction with parameterized many body
calculations based on a single model Hamiltonian with the same set
of parameter values for both core level and valence band spectra.
Here, we also study the evolution of the core level and valence
band spectra across the solid solution and discuss their
implications.

\section{EXPERIMENTAL AND THEORETICAL DETAILS}

Samples of NiS$_{2-x}$Se$_x$ with $x=0.0$, 0.4, 0.6, 0.8 and 1.2
used  for  the present  study  were prepared  by the standard
solid state reaction  techniques reported in the literature
\cite{ips-prb}.   X-ray diffraction patterns as well as
resistivities of the samples were found to be in agreement with
the reported data \cite{wil@4,miyadai@4}. Spectroscopic
measurements   were carried   out  in  a   combined  VSW
spectrometer  with a  base pressure  of 2$\times10^{-10}$  mbar
equipped with a monochromatized Al~K$\alpha$  x-ray source  with an
overall instrumental resolution better than  0.8~eV.   All the
experiments were performed  at 120~K and the sample  surface was
cleaned {\it in  situ} periodically during  the experiments by
scraping  with an alumina file; the surface  cleanliness was
monitored by  recording the carbon 1$s$ and oxygen 1$s$  core
level XP signals.  The reproducibility of the spectral features
were confirmed in each case.  The binding energy was  calibrated
to the  instrument Fermi-level  that was  determined by recording
the Fermi-edge region of a clean silver  sample.

Core  level and valence  band~(VB) spectra  were calculated  for
NiS$_6$ cluster with  an octahedral structure, within  a
parameterized many-body multi-band   model   including   orbital
dependent   electron-electron (multiplet)  interactions; the
details of these calculations have been described elsewhere
\cite{Dimen@4,nis-prb,BaCo@4,millPRB}. The calculations  were
performed in the symmetry adapted  $t_{2g}$ and $e_g$  basis
including  the TM 3$d$ and  the  bonding  sulphur  3$p$ orbitals.
In the calculation for  the valence band spectrum, the S~3$p$
spectral contribution was obtained by the resolution  broadening
of the S~3$p$ partial DOS obtained from  the LMTO band  structure
calculation \cite{ips-prb}, while the Ni~3$d$ contribution to  the
spectrum was evaluated within the cluster model.  This is
reasonable in view of the negligible correlation  effects within
the broad band S~3$p$ manifold.   Moreover,  the  S~3$p$  spectral
distribution is  strongly influenced by the short S-S bonds in the
S$^{2-}_2$ dimers, not included in  the  cluster  model.    In
contrast, such  effects  are  described accurately within the LMTO
approach. The calculations were performed by the Lanczos algorithm
and  the calculated one-electron  removal spectra were
appropriately broadened to simulate the  experimental spectra.  In
the Ni~2$p$ core level calculation,
Doniach-$\breve{S}$unji$\acute{c}$ line shape function \cite{DS}
was used for  broadening the  discrete energy  spectrum of  the
cluster model, in order  to represent the asymmetric line shape of
core levels and is also consistent with other core levels in the
system (\eg S~2$p$ and Se~3$d$).  In  the case of valence band
spectral calculations,  energy dependent Lorentzian function was
used for the lifetime broadening. Other broadening effects arising
from the resolution broadening and solid  state effects, such as
the band structure and phonon broadenings, were taken into account
by convoluting the spectra with a Gaussian function. The
broadening parameters were  found to be consistent with values
used earlier for similar systems
\cite{Dimen@4,nis-prb,BaCo@4,millPRB}.

\section{RESULTS \& DISCUSSIONS}

The S~2$p$  core level spectra for the samples studied are shown
in the main panel  of Fig.~1,  as open  circles.  For  NiS$_2$,
the experimental S~2$p$ spectrum shows a spin-orbit  split doublet
as expected and the overlapping solid line shows the simulated
spectrum using the usual constrain on the intensity ratio (2:1)
between the spin-orbit split partners, 2$p_{3/2}$ and
2$p_{1/2}$~\cite{jagjit}. However, for $x>0$, we see two more
features in the spectra, indicating the presence of overlapping
Se~3$p$ levels in the same binding energy range. We have analyzed
the experimental spectra for $x > 0$ samples in terms of
contribution of two spin-orbit split doublets, simulating the
S~2$p$ and Se~3$p$ states  and the resulting fits from
least-square-error analysis are shown as  the overlapping solid
lines in each case, illustrating a very good agreement.  In the
inset~I, we show the individual components of the S~2$p$ and
Se~3$p$, separately for $x=0.8$. In order  to estimate the
relative Se/S ratio  in the surface region of  the samples probed
by the photoemission technique, we take  the ratio of the
intensities of Se~3$p$ and S~2$p$ for $x = 0.4$,  0.6 and 1.2 and
normalize with that obtained for $x = 0.8$, assuming that for $x =
0.8$, the surface composition is the same as dictated by the
stoichiometry. Thus obtained ratio
$\frac{(Se~3p/S~2p)_x}{(Se~3p/S~2p)_{0.8}}$ from the fitting
procedure are plotted in inset~II as open circles with error bars.
We also plot the expected ratio (solid line) in the same inset as
a  function of the nominal bulk compositions for all $x>0$.  The
plot exhibits a good agreement between the experimental and the
expected values. This clearly suggests that  the surface
composition  in this system remains close to that of the bulk
without any complication arising from surface  non-stoichiometry
or segregation.

The Se~3$d$  spectra for  the entire composition  range is shown
in the main panel  of Fig.~2.  The  spectral features are broad
without any clear   indication  of  the expected  spin-orbit split
doublet   ($3d_{5/2}$  and $3d_{3/2}$) signals. Our analysis of
the spectral  line shape   suggests it to be incompatible  with a
single type  of  Se atoms  in the system, since the spectral shape
could not be simulated by a single set of spin-orbit doublet.  It
is reasonable to expect the presence of  two types of Se atoms in
NiS$_{2-x}$Se$_x$.  These two types of Se atoms are distinguished
by the bonded partner within the dimer  unit,   since  one   would
in  general   expect (Se-Se)$^{2-}$, (Se-S)$^{2-}$ as well as
(S-S)$^{2-}$ units to be present in samples with $2>x>0$ in the
NiS$_{2-x}$Se$_x$  series. Raman spectroscopy \cite{raman@4} has
been used to  identify the presence of bond  stretching vibrations
of S-S, S-Se  and Se-Se molecular  units  in agreement with this
point of view. However, the relative abundance of each of these
three types of dimers in any sample of given $x$ is not known so
far.    Since  the nearest neighbor chemical environment  of Se in
(Se-Se)$^{2-}$ is different from  Se in (Se-S)$^{2-}$ dimers, the
two types of Se are expected to have different binding energies
arising from chemical shifts in the core level spectra. Thus, we
attempt to describe the spectra  with two distinct Se~3$d$
spin-orbit split doublets.  In  inset~I,  the two components
obtained  for $x=1.2$  along with  the experimental data and the
resulting fit are shown as an  example.  The resulting  fits to
the experimental  spectra are shown  as solid lines  in the main
panel of Fig.~2, exhibiting  good agreement for all $x$.  The
observed chemical shift of about  1.0~eV between the two types of
Se sites, was found to be the same  in all  the samples. It  is
easy  to estimate the  expected intensity ratio from these two
types of Se sites assuming a random substitution of S atoms by Se
in   NiS$_{2-x}$Se$_x$;    this is    given   by    the
statistical ratio, $\frac{I_{Se-Se}}{I_{Se-S}}=\frac{x}{4(2-x)}$.
In inset~II, we plot experimentally and theoretically obtained
intensity ratios between Se-Se and Se-S pairs (open circles and
solid line respectively) against the respective Se content ($x$).
We find a  remarkable agreement between the two, indicating  that
the Se substitution is indeed random in these samples.

We now turn to the Ni~2$p$ core level spectrum which often
manifests distinct spectral signatures arising from various
many-body interactions. Ni~2$p$   spectrum in NiS$_2$ (see Fig.~3)
consists  of  spin-orbit  split,  $2p_{3/2}$  and  $2p_{1/2}$
peaks  at 853.5~eV  and  871~eV binding  energies,  respectively,
with  pronounced satellite features around 860~eV  and 876~eV,
indicating the presence of electron correlations  in the system.
The satellite intensity relative to  the main peak appears
considerably more intense  in the  $2p_{1/2}$ region  compared to
that in  the $2p_{3/2}$ region.   In order to determine the
inelastic scattering background, we have  performed  electron
energy  loss  spectroscopy  (EELS)  on  these samples, with the
same primary energy  as that of the Ni~$2p$ core level peak. Using
a procedure that have been previously employed \cite{nis-prb,BaCo@4,millPRB},
the inelastic background  function obtained for NiS$_2$ is shown
in the inset of Fig.~3,  as  a  dotted line.   We  find  that
there  is an  intense  and structured  contribution from  the
background  function  overlapping the $2p_{1/2}$  satellite
region,  resulting  in   the  anomalously  large satellite
intensity  in the  $2p_{1/2}$ region compared  to that  in the
$2p_{3/2}$ region; this feature  in the inelastic scattering
spectrum of NiS$_2$  arises from  a plasmon  band.  It  is also
seen that  at about 857~eV, there is a peak-like structure in the
inelastic background; this appears  at  about  the same  energy
position  as  that of  the  strong asymmetry in the line shape of
the $2p_{3/2}$ main peak.  This structure in the  inelastic
scattering background  could have its origin  from the inter-band
$p-d$ transitions.

We  have calculated Ni~$2p$ core level and valence  band (VB)
spectra of NiS$_2$ within the same  model involving  a NiS$_6$
cluster  to  obtain quantitative  many-body description  of the
electronic structure. In this calculation for   Ni$^{2+}$,   the
electron-electron   interaction parameters, $F^2_{dd}  = 9.79$~eV,
$F^4_{dd} = 6.08$~eV, $F^2_{pd} = 6.68$~eV, $G^1_{pd}  = 5.07$~eV,
and $G^3_{pd} = 2.88$~eV were used. The  calculated   Ni~$2p$
spectrum  with the hopping interaction strength $(pd\sigma)$, the
charge transfer energy ($\Delta$) and Coulomb interaction
strength ($U_{dd}$) being
-1.5~eV, 2.0~eV and 4.0~eV, respectively, is shown in
the main figure by  a solid line
overlapping  the experimental spectrum  (open circles).   The
calculated spectrum includes the experimentally determined
inelastic background, shown in  the inset.   There is evidently  a
good agreement between the experimental  and  the  calculated
spectrum.   The calculated discrete spectrum arising from this
finite-sized cluster calculation without any broadening is also
presented as a stick diagram in  the main panel.  The present
results show  that it is necessary  to take  into  account the
contributions  to the experimental  spectrum  from  the extrinsic
loss processes in order  to provide a proper quantitative
description of the spectrum.   The  previous estimates  of
various  parameter strengths  in NiS$_2$ obtained  from a  model
that included  a ``conduction  band'' in addition to  the
Ni~3$d$-S~3$p$ basis within the cluster model  for the core level
calculation \cite{fujicorecal@4}, are $(pd\sigma) = -1.2$~eV,
$\Delta = 2.0$~eV, and $U_{dd} = 5.5$~eV.  Thus, the present
estimates differ significantly for both $(pd\sigma)$ and $U_{dd}$,
where we have a larger estimate for $(pd\sigma)$  and  a  smaller
value for the  $U_{dd}$.   However,  the $(pd\sigma)$ values
estimated for NiS$_2$ in the present case is similar to  that
estimated  for other  divalent nickel  sulphides,  for example,
$(pd\sigma)   = -1.4$~eV   for   NiS \cite{nis-prb}  and   -1.5~eV
for BaNiS$_2$ \cite{BaCo@4}.   Additionally, as we  show later  in the  text, the
present estimates  are also consistent  with the valence  band spectrum.
The charge transfer energy,  $\Delta$, varies considerably for different
sulphides of  nickel, with NiS$_2$  having a $\Delta$ of 2~eV  compared to
2.5~eV for NiS and 1.0~eV for BaNiS$_2$  \cite{BaCo@4}.  In general,
$\Delta$ is  expected to  be smaller for  sulphides compared  to oxides,
since the  O~$2p$ levels are  energetically more stable than  the S~$3p$
levels;  for   example,  the  estimated  $\Delta$  for   NiO  is  5.5~eV
 \cite{Dimen@4}.  The value of $U_{dd}$ in  NiS$_2$ is found to be the same
as that  in NiS \cite{nis-prb}, while in  the case
of BaNiS$_2$, $U_{dd}$ estimated is still  smaller ($\sim 3$~eV)
\cite{BaCo@4}, possibly arising from a more efficient screening in the metallic system due to a smaller $\Delta$ and slightly large $(pd\sigma)$ values.

The   ground  state   wavefunction  of NiS$_2$ corresponding to
the  estimated  parameter strengths have  been analyzed in terms
of  contributions from various  electron configurations.
The ground state  of the  system was found   to   consist   of 61.6\%,
35.1\%   and   3.3\%   of   $d^8$, $d^9\underline{L}^1$, and
$d^{10}\underline{L}^2$ configurations with a high-spin
configuration ($S=1$).  The average value of the $d-$occupancy
$(n_d)$ is found  to be 8.42, showing a highly  covalent ground
state of the system,  which is very similar  to that obtained  for
NiS (8.43) \cite{nis-prb} and BaNiS$_2$ (8.48) \cite{BaCo@4}.
We have analyzed the characters of  the final states of the system
responsible for the different features in the experimental
spectrum in order to understand their origins.  The analysis was
carried out for some of the representative final state energies
marked 1-10  in Fig.~3.  The different contributions to  the final
states from  various electron configurations ($d^8$,
$d^9\underline{L}^1$,  $d^{10}\underline{L}^2$)  are listed  in
Table.~I.  These features can be grouped into three different
regions; the  main  peak  region,  852-856~eV (labelled  1-4);
intense  satellite region, 859-862~eV (labelled  5-8), and weak
satellites in  the region of 864-867~eV (labelled 9 and 10).  The
first group of features in the main peak region  has a dominant
$d^9\underline{L}^1$ character  as seen from the  table,  which
are   the  `well-screened'  states  of  the  system, corresponding
to  one ligand (sulphur) electron being  transferred to the Ni
site  to screen  the attractive potential of the  Ni~$2p$
core-hole created by the photoemission  process.   This is similar
to the observations  from previous    studies   in the charge
transfer    systems,   where $d^9\underline{L}^1$ states are
stabilized  compared  to  the  other configurations giving rise to
the intense main peak.   The second group of features have a mixed
character with  significant contributions from all the
configurations.  This is in  contrast to the case  of NiO where
the  intense  satellite  structure  results  primarily from the
$d^8$ configuration, establishing that the satellite in the Ni
2$p$ core spectrum in NiS$_2$ cannot be described as a 'poorly
screened' state. Such heavily mixed characters of the satellites
have been shown to exist for intermetallic compounds of Th
\cite{Intermetallics}. For the third group   of features, the
primary contribution comes from both $d^8$   and
$d^{9}\underline{L}^1$ contributions with relatively lower
contributions   coming from $d^{10}\underline{L}^2$ character.

In Fig.~4, we show the experimental XP valence band spectrum (open
circles) along with the calculated spectrum (solid line) using the
same model. As  mentioned before, we have used the S~3$p$  partial
DOS  obtained  from the  band  structure calculation  to represent
the S~3$p$ contribution  to the valence  band spectrum.   It was
also found necessary  to shift rigidly the S~3$p$  partial DOS by
about 0.9~eV to  higher binding energy in order to match the
experimentally observed S~3$p$  features.  The  various
contributions to the calculated spectrum, Ni~$3d$ (dashed  line)
and S~$3p$ (dot-dash line) are shown along with the  experimental
data in Fig.~4.  An inelastic  scattering background function
(dotted lines)  is also included  in the total calculated
spectrum.  The calculated discrete contributions from the Ni~$3d$
to the total spectrum for  NiS$_2$ are also  shown without  any
broadening effects as  a stick diagram in  Fig.~4.  The  parameter
set used for the valence band  calculation is identical to  that
used for the core level calculation. In  view of the fact that  no
parameter was adjusted to  obtain a fit,  the agreement between
the experimental spectrum and the calculated  one is remarkable
over the  entire  energy   range.   The  increasing  intensity  in
the experimental spectrum beyond 11~eV is due to S~3$s$ level with
a peak at about 14~eV.   There are two distinct sets of parameters
proposed earlier on the basis  of valence band analysis.  Fujimori
\etal \cite{fujivbcal@4}  obtained $\Delta = 1.8$~eV,   $U_{dd}  =
3.3$~eV  and   $(pd\sigma)  =   -1.5$~eV, while Sangaletti \etal
\cite{luigi@4}  arrived  at  $\Delta  = 3.0$~eV, $U_{dd}  =
4.5$~eV  and $(pd\sigma)   =  -1.35$~eV. Good agreement
between  the experimental spectrum and  the calculated one in the
present study (see Fig.~4)  over the  entire range with  a minimum
number of  parameters, indicates the reliability of the  parameter
set estimated here. This is further  enhanced by the  fact that
the  same set  of parameters  also provides an equally
satisfactory  description of the core level spectrum (see Fig.~3).
We note that previous estimates in ref.~\cite{fujivbcal@4} is in
better agreement with the present results, with the
ref.~\cite{luigi@4} arriving at a too high an estimate for $\Delta$
and too low an estimate for $(pd\sigma)$.

The main peak region in the valance band spectrum at about 2.3~eV
arises essentially from  Ni~3$d$ photoemission  contribution
though there  is a small contribution arising from  S~3$p$ states
also due to hybridization mixing of Ni~3$d$ and S~3$p$  states.
The features at 3.5~eV and 7.5~eV are  contributed primarily  by
the  S~3$p$ contributions.   As  the same model  and same
parameters were used for the  VB calculation of NiS$_2$, the
ground  state of the system was the same as that described  in the
core level calculation.  The  results of the character analysis of
the final states  labelled 1-11  in Fig.~4  are shown  in
Table.~II.   The spectral features  can  be grouped  into  three
regions,  the main  peak  region (0-3.5~eV,  labelled 1-4),  the
spectral features  in  the 5-7~eV  range (labelled  5-7) and
satellites  beyond 8.5~eV  (marked 8-11).   The final states   in
the   main    peak   region   predominantly   consist   of
$d^8\underline{L}^1$  states   with  non-negligible  contributions
from $d^9\underline{L}^2$ and  $d^7$ configurations.  This is
similar to the case of  other charge-transfer  systems, like
NiO~\cite{Dimen@4}. In the 5-7~eV spectral region, the final
states have very similar character as that in the  main peak
region, with the contributions from $d^8\underline{L}^1$ further
enhanced at the expense of contributions from $d^7$ and
$d^{10}\underline{L}^3$ states. The satellite  features at higher
energies  (marked 8-11) are dominated  by $d^7$  and
$d^9\underline{L}^2$ configurations with little contributions from
other configurations. As these satellite features have dominant
$d^7$ character, these could be attributed to the spectral
signature of the lower Hubbard band in the system. However, these
features are not distinct in the experimental spectrum due to
their weak intensities.

We now turn to the results obtained from the solid solution
NiS$_{2-x}$Se$_x$ in order to address the changing electronic
structure observed with increasing $x$ in the series. The  Ni~$2p$
core level  spectra  for the  entire series  are  shown  in the
main panel of Fig.~5. For all the compositions the spectra appear
to be quite similar, though there are some subtle differences
between the spectra with different Se contents. The spectra  for
$x=0$, 0.6 and 1.2 are overlapped in the inset of Fig.~5 for the
2$p_{3/2}$ region. As the Se content increases, the 2$p_{3/2}$
level narrows, consistent with the previous report \cite{haas@4}.
In this comparison, the satellite intensity relative to the main
peak intensity appears to decrease marginally as $x$ increases.
This apparent decrease of the satellite intensity is essentially
compensated by the narrowing of the main peak, such that the
integrated satellite intensity relative to the main peak
integrated intensity remains essentially the same. From the core
level analysis, we see that the intensity of the satellite peak is
sensitive to the value of $U$. Hence, on the basis of the
insensitivity of the satellite intensity, we conclude that all
electronic interaction strengths, and in particular the on-site
coulomb interaction strength $U$, do not change significantly
across the composition range studied; the same calculated result,
as shown in Fig.~3, can explain the different core level spectra
in Fig.~5 equally well with slight adjustments of the broadening
functions.

The XP valence band spectra of  NiS$_{2-x}$Se$_x$ for  $x = 0.0$,
0.6, 0.8 and 1.2 recorded using Al K$\alpha$ are shown  in Fig.~6.
As $x$ increases, the various features marked  (A, B, and C)
become more evident in the  XP spectra; additionally, the
separation between the features A and B increases  across the
series.   As $x$ increases, the Se~4$p$ contribution to the VB
increases and the changes in feature  C can be attributed to this.
As the features A and B are dominated by Ni~3$d$ states, the
spectral changes suggest some subtle modifications in the
electronic structure, which is presumably also responsible for the
change in the ground state properties with $x$, namely the
insulator-metal transition. However, such effects cannot be
treated within the minimal cluster model considered here and more
sophisticated approaches like Dynamical Mean Field Theory (DMFT)
\cite{shen2@4,watanabe@4} are required to study the detailed
electronic structure near $E_F$. Within the resolution limit of
XPS, we do not see any dramatic changes in the valence band
spectrum near $E_F$, across the series. Even for the bulk
insulating NiS$_2$, there is a finite intensity at $E_F$ and this
could be due to the resolution broadening, but also might have
some contribution coming from the surface metallic layer
\cite{nis2prl}. However, there is an indication of an increased
intensity at $E_F$ with increasing $x$, consistent with the
increase in the metallicity of the solid solution. The
metal-insulator transition in NiS$_{2-x}$Se$_x$ series is an issue
which have been studied extensively, however, still not understood
completely. Our study reveals that the on-site Coulomb interaction
in the system  does not change with the increase in the Se
substitution. This is consistent with the suggestion of NiS$_2$
and NiSe$_2$ having similar $U$ \cite{ips-prb}. Moreover, band
structure  studies \cite{ips-prb} reveal that the effective Ni~$d$
band width ($W$) increases in going from NiS$_2$ to NiSe$_2$.
Thus, as a result of the increase in $W$, the effective
correlation strength ($U/W$) decreases, driving the system
metallic for $x$ above $x_c$. The estimated values of
$(pd\sigma)$, $\Delta$ and $U$ (-1.5~eV, 2.0~eV and 4.0~eV,
respectively), places NiS$_2$ in the regime of covalent
insulators,\cite{seva} close to $pd$-metals. As Se is substituted
in place of S, the system moves in to the $pd$-metallic regime,
driven by the decrease in $(U/W)$, resulting in the bulk
metal-insulator transition in the system.

In  conclusion, we have investigated the electronic structure of
NiS$_{2-x}$Se$_x$ system using x-ray photoemission spectroscopy.
The analysis of the S~2$p$ and Se~3$d$ spectra revealed the
homogeneity of the samples, without any segregation or
non-stoichiometry. The electronic structure of NiS$_2$ has been
studied by means of a parameterized multi-band cluster model and
is found to be successful in describing the core level and valence
band spectra within the same model and an identical parameter set.
These calculations show that NiS$_2$ is a strongly correlated
system with a highly covalent character. It is found that the
on-site Coulomb interaction strength ($U$) does not change with
the Se substitution  and the system transforms from the covalent
insulator regime to a $pd$-metallic type, due to the enhanced
bandwidth resulting from the substitution of S by Se.

\section{Acknowledgments}

The authors thank the Department of Science and Technology, and
the Board of Research in Nuclear Sciences, Government of India,
for financial support.  SRK  thanks the Council of Scientific and
Industrial Research, Government of India, and The Abdus Salam
International Centre for Theoretical Physics (ICTP), Trieste,
Italy for financial assistances. The authors also thank Professor
S.  Ramasesha and the Supercomputer Education and Research Center,
Indian Institute of Science, for providing the computational
facility.

\pagebreak

\section{figure captions}

Fig.~1.  S~2$p$ core level spectral region for the series
NiS$_{2-x}$Se$_x$. For $x > 0$, Se~3$p$ contribution in the same
spectral region is observed. The solid lines show the result of
the analysis of the spectral shape in terms of contributions from
S~2$p$ and Se~3$p$ levels. Individual  S~2$p$ and Se~3$p$
components  are shown for $x = 0.8$ in inset~I. The Se/S ratio
with respect to $x = 0.8$ obtained from the analysis  is plotted
in inset~II as a function of  the nominal Se content of the
samples.

Fig.~2. Se~3$d$ core level spectra for the series
NiS$_{2-x}$Se$_x$. The solid lines show the result of the analysis
of the spectral shape in terms of contributions from  two
chemically-distinct  Se~3$d$ components arising from Se-Se and
Se-S pairs. The spectral analysis is illustrated for $x = 1.2$ in
inset~I in terms of the Se~3$d$ components. The  intensity ratio
between different components (Se-Se and Se-S bonds) obtained from
the analysis is shown in inset~II as open circles while the solid
line represents theoretically expected ratio.

Fig.~3. Experimental Ni~$2p$ core level spectrum of NiS$_2$ (solid
circles) along with the inelastic scattering background function
(dotted line) obtained from EELS is shown in the inset.
Experimental Ni~$2p$ spectrum (open circles) along with the
calculated spectrum (solid line) for NiS$_2$ obtained from the
cluster calculation is shown in the main panel.  Various final
states of the cluster calculation and the corresponding intensity
contributions without any broadening are shown as the bar diagram.

Fig.~4. The experimental VB spectrum (open circles) along with the
calculated spectrum (solid line), Ni~$3d$ component (dashed line),
S~$3p$ component (dot-dashed line) and the integral background
(dotted line) are shown for NiS$_2$. The final states of the
calculation and the corresponding intensities without any
broadening are shown as the bar diagram.

Fig.~5. Ni~2$p$ core level spectra from the series
NiS$_{2-x}$Se$_x$ for various $x$ values are shown in the main
panel. Inset shows the Ni~2$p_{3/2}$ region for $x=0$, 0.6 and 1.2.

Fig.~6. Valence band spectra obtained using Al~K$\alpha$
radiation for NiS$_{2-x}$Se$_x$. Various spectral features are
marked A, B and C and their evolution across the series is
discussed in the text.

\onecolumn

\begin{table}
TABLE  I.~~~Contributions from various   configurations  in the  final
states of the Ni~2$p$ core level photoemission in NiS$_2$.\\
\begin{tabular}{c|c c c c c c c c c c}
Peak no. & 1 & 2 & 3 & 4 & 5 & 6 & 7 & 8 & 9 & 10 \\ BE & 853.6
&  854.3  & 854.9  & 855.6  & 859.5 & 860.2 & 861.4 & 861.8 &
 864.5 & 866.6\\ \hline

$d^8$ & 26.63 & 16.16 & 21.24 & 9.51 &
40.92 & 37.49 & 34.40 & 21.66 & 35.03 & 58.34  \\

$d^9\underline{L}^1$ & 56.49
& 61.47 & 60.41 & 63.89 &  29.06 & 25.33  & 12.88 & 23.09 &
41.15 & 32.45  \\

$d^{10}\underline{L}^2$  & 16.88 &  22.37 &
18.35 & 26.60 &  30.02 & 37.18 & 52.72 & 55.25 & 23.82 & 9.21 \\
\end{tabular}
\end{table}

\begin{table}
TABLE II.~~~Contributions from  various  configurations  in the  final
states of valence-band photoemission in NiS$_2$.\\
\begin{tabular}{c|c c c c c c c c c c c}
Peak no. & 1 & 2 & 3 & 4 & 5 & 6 & 7 & 8 & 9 &  10 & 11\\

BE & 2.1 & 2.3&  3.0 & 3.3 & 5.4  & 5.9 & 6.7 &  8.7 & 9.1 &
9.7 & 10.0\\ \hline

$d^7$ & 14.50 & 23.86 & 14.75 & 13.71 & 3.44 & 0.00  & 0.38 & 49.33 &
35.43 & 30.06 & 44.72\\

$d^8\underline{L}^1$ & 54.53 & 54.34 & 56.63 & 56.13 & 68.17 & 76.32 &
67.05 & 7.15 & 11.32 & 11.06 & 4.65 \\

$d^9\underline{L}^2$ & 28.26 & 20.38 & 26.52 & 27.92 & 27.24 & 23.68 &
32.12 & 37.25 & 42.77 & 52.17 & 41.59 \\

$d^{10}\underline{L}^3$ & 2.71 & 1.42 & 2.10  & 2.24 & 1.15 &
0.00 & 0.45 & 6.27 & 10.48 & 6.71 & 9.04 \\
\end{tabular}
\end{table}

\end{document}